\documentclass{jaa}
\usepackage{natbib}
\usepackage{xcolor}

\DeclareRobustCommand{\VAN}[3]{#2}
\let\VANthebibliography\thebibliography
\def\thebibliography{\DeclareRobustCommand{\VAN}[3]{##3}\VANthebibliography}

\usepackage{graphicx}	
\usepackage{amsmath}	
\usepackage{amssymb}	
\usepackage{array, multirow}
\usepackage{tabularx}
\usepackage{caption}
\usepackage[colorlinks=true,citecolor=blue,urlcolor=blue]{hyperref}
\usepackage{booktabs}
\usepackage{ltablex}
\usepackage{aas_macros}

\begin{document}
\sloppy

\title{\czti\ searches for hard X-ray prompt emission from Fast Radio Bursts}

\newcommand\aastex{AAS\TeX}
\newcommand\latex{La\TeX}
\newcommand{\vb}[1]{{\color{purple} #1}}
\newcommand{\del}[1]{\textcolor{gray}{#1}}
\newcommand{\vect}[1]{\ensuremath{\boldsymbol{#1}}}
\newcommand{\rem}[1]{\textcolor{red}{\textbf{--- #1 ---}}}
\newcommand{\outline}[1]{\textit{#1}}
\newcommand{\rough}[1]{\textit{\color{brown} #1}}
\newcommand{\todo}[1]{\textbf{\color{red} #1}}
\newcommand{\asat}{{\em AstroSat}}
\newcommand{\fermi}{{\em Fermi}}
\newcommand{\swift}{{\em Swift}}
\newcommand{\svom}{{\em SVOM}}
\newcommand{\daksha}{{\em Daksha}}
\newcommand{\czti}{{\asat-CZTI}}
\newcommand{\tnt}{\ensuremath{\mathrm{T}_{90}}}
\newcommand{\ch}[1]{{\rough{#1}}}
\newcommand{\gfo}{GW170817}
\newcommand{\ifh}[1]{{\bigskip \rough{Insert figure here: #1} \\}}
\newcommand{\egs}{$\mathrm{ergs~cm^{-2}~s^{-1}}$}
\newcommand{\eg}{$\mathrm{ergs~cm^{-2}}$}
\newcommand{\es}{$\mathrm{ergs~s^{-1}}$}
\newcommand{\dt}{detector}
\newcommand{\dts}{detectors}
\newcommand{\msun}{\ensuremath{\mathrm{M}_\odot}}
\newcommand{\joss}{Journal of Open Source Software}
\newcommand{\spie}{Society of Photo-Optical Instrumentation Engineers (SPIE) Conference Series}
\newcommand{\degr}{\ensuremath{^\circ}}


\author{\href{https://orcid.org/0000-0003-3630-9440}{G. Waratkar}\textsuperscript{1,*}, \href{https://orcid.org/0000-0002-6429-8139}{M. Dixit}\textsuperscript{2}, \href{https://orcid.org/0000-0003-3630-9440}{S. P. Tendulkar}\textsuperscript{3}, \href{https://orcid.org/0000-0002-6112-7609}{V. Bhalerao}\textsuperscript{1}, \href{http://orcid.org/0000-0003-3352-3142}{D. Bhattacharya}\textsuperscript{4,5}, \href{https://orcid.org/0000-0002-2050-0913}{S. Vadawale}\textsuperscript{6}}

\affilOne{\textsuperscript{1}Department of Physics, Indian Institute of Technology Bombay, Powai, 400 076, India.\\}
\affilTwo{\textsuperscript{2}Department of Aerospace, Indian Institute of Technology Bombay, Powai, 400 076, India.\\}
\affilThree{\textsuperscript{3}Tata Institute of Fundamental Research, Homi Bhabha Road, Mumbai 400005, India.\\}
\affilFour{\textsuperscript{4}Ashoka University, Department of Physics, Sonepat, Haryana 131029, India.\\}
\affilFive{\textsuperscript{5}Inter-University Center for Astronomy and Astrophysics, Pune, Maharashtra-411007, India.\\}
\affilSix{\textsuperscript{6}Physical Research Laboratory, Ahmedabad 380009, India.\\}


\twocolumn[{

\maketitle

\corres{gauravwaratkar@iitb.ac.in}

\msinfo{1 March 2025}{1 March 2025}

\begin{abstract}
Fast Radio Bursts (FRBs) are short-duration, highly-energetic extragalactic radio transients with unclear origins \& emission mechanisms. Despite extensive multi-wavelength searches, no credible X-ray or other prompt electromagnetic counterparts have been found for extragalactic FRBs. We present results from a comprehensive search for such prompt X-ray counterparts using \czti\, which has been actively detecting other high-energy fast transients like Gamma-ray bursts (GRBs). We undertook a systematic search in \czti\ data for hard X-ray transients temporally \& spatially coincident with 578 FRBs, and found no X-ray counterparts. We estimate flux upper limits for these events and convert them to upper limits on X-ray-to-radio fluence ratios. Further, we utilize the redshifts derived from the dispersion measures of these FRBs to compare their isotropic luminosities with those of GRBs, providing insights into potential similarities between these two classes of transients. Finally, we explore the prospects for X-ray counterpart detections using other current and upcoming X-ray monitors, including \fermi-GBM, \swift-BAT, \svom-ECLAIRs, and \daksha, in the era of next-generation FRB detection facilities such as CHIME, DSA-2000, CHORD, and BURSTT. Our results highlight that highly sensitive X-ray monitors with large sky coverage, like \daksha, will provide the best opportunities to detect X-ray counterparts of bright FRBs.  
\end{abstract}


}]


\doinum{12.3456/s78910-011-012-3}
\artcitid{\#\#\#\#}
\volnum{000}
\year{0000}
\pgrange{1--}
\setcounter{page}{1}
\lp{13}

\sloppy
\section{Introduction}
Fast Radio Bursts (FRBs) are millisecond-duration radio pulses of extragalactic origin, first identified in 2007 \citep{lorimer2007}. Despite the detection of over a thousand FRB bursts to date \citep{chimecat12021}, their nature remains a mystery. Numerous theoretical models have been proposed --- ranging from cataclysmic events to acitivity in persistent astrophysical sources --- but no single or combination of models explain the variety of observations \citep{frbtheorycat2019}\footnote{\url{https://frbtheorycat.org/}}. While FRBs have been mainly studied in the radio band, covering frequencies from 110~MHz to 8~GHz, their mysterious nature has prompted searches across the electromagnetic spectrum, providing important insights and constraints \citep{cunningham2019,hxmt2020,akash2020,agile2021,lat2022,konus2024}. 

Several models predict high-energy emission associated with FRBs, while others do not \citep{frbtheorycat2019} and the detection of a high-energy counterpart would be a great discriminator. Some of these models compare the mechanisms behind FRBs to those of gamma-ray bursts (GRBs), particularly synchrotron radiation from relativistic shock collisions \citep{zhangnat2020,lyubarsky2021}. Further, these theoretical models expect that the radio emission is a very small fraction of the total energetics and hence the high-energy emission could be much brighter \citep{zhangnat2020,sgrimplication2020,petroff2022}. This suggested connection between FRBs and GRBs has led to extensive searches for coincidences between FRBs, GRBs, and gravitational wave mergers \citep{gwfrb2019,gwfrb2023,grbfrb2023,gwmagnetar2024,gwfrb2024,frbgrb2024}. A proposed link between the binary neutron star merger GW190425 and FRB20190425A \citep{frbbns2023} is still being debated \citep{bnsfrb2024}. 

In April 2020, the Galactic magnetar SGR~1935+2154 emitted a bright, millisecond-duration radio burst, similar to extragalactic FRBs, which was observed by multiple radio observatories. This was the first radio detection of such a burst from a Galactic source, further supporting the connection between FRBs and magnetars \citep{sgr12020, sgr22020}. At the same time, several X-ray bursts were detected by missions like NICER, INTEGRAL, the \emph{Insight}-HXMT, and \fermi-GBM, showing the high-energy nature of these emissions \citep{nicersgr2020,integralsgr2020,hxmtsgr2021,gbmsgr2020}. These simultaneous detections have strengthened the idea that magnetars might be the source of at least some FRBs \citep{sgrimplication2020}.These observations further highlight the need to search for high-energy counterparts of extragalactic FRBs. Such searches could help determine whether all FRBs have high-energy emissions or if only some do, providing more understanding of the environments and physical processes involved. 

X-ray searches for prompt counterparts to extragalactic FRBs have been conducted using various instruments, including those in dedicated simultaneous search programs like Deeper, Wider, Faster \citep[DWF; ][]{dwf2019}, as well as independent searches by several hard X-ray observatories such as \emph{INSIGHT}-HXMT \citep{hxmt2020}, \fermi-GBM \citep{gbmfrb2019}, \swift-BAT \citep{cunningham2019}, \emph{AGILE} \citep{agile2021}, and KONUS-Wind \citep{konus2024}. The observed detection rate of X-ray transients, like GRBs, is much lower than the detection rate of FRBs, suggesting that most FRBs will not have detectable high-energy counterpart \citep{kienlingbmcat2020,lienbatcatalog2016,chimecat12021,zhangfast2024}. While these searches have yielded no credible counterparts, they have provided key insights into the limits of high-energy emission from FRBs. The absence of detections has led to stringent upper limits on the X-ray emission from FRBs, which constrain the energetics and emission mechanisms of these transients \citep{cunningham2019} which highlights the need to keep searching for these elusive counterparts. 

\asat\ is an Indian space telescope launched in September 2015 \citep{astrosat2014}. The Cadmium Zinc Telluride Imager \citep[CZTI; ][]{czti2017} onboard \asat\ is designed as a hard X-ray instrument operating in the energy range of 20--200keV, with an on-axis field of view (FOV) of approximately $4.6\degr \times 4.6\degr$. Beyond its primary observational capabilities, CZTI functions as an all-sky monitor with a wide angular response, enabling the detection of transient events across the sky \citep{cift2021,waratkar2024}. Since its launch, CZTI has actively monitored gamma-ray bursts (GRBs), detecting over 650 GRBs\footnote{\url{https://astrosat.iucaa.in/czti/grb}}. While its localization capabilities are limited to select bright high-energy transients \citep{cztiloc2024}, CZTI has been regularly utilized as a node in the InterPlanetary Network to improve the localization of other interesting transients \citep{ipngcn1,ipngcn2}. Owing to these capabilities, CZTI has been previously employed in the search for counterparts to gravitational wave mergers \citep{waratkar2024} and FRBs \citep{akash2020}. The prior search for FRB counterparts resulted in no detections but provided highly competitive upper limits on X-ray emission relative to contemporaneous instruments \citep{akash2020}.

Since our previous work, the catalog of known FRBs has expanded by an order of magnitude, primarily due to the detections by CHIME \citep{chimecat12021}. Additionally, advancements in data processing techniques for CZTI, as outlined in \citet{ratheesh2021, waratkar2024}, have enhanced our ability to detect fainter signals and set more stringent upper limits. In this study, we revisit the search for high-energy counterparts to FRBs using the updated methodology and an expanded dataset. In Section~\ref{sec:sample}, we detail the FRB sample selection criteria. Section~\ref{sec:dataanalysis} outlines the calculations of FRB arrival times corresponding to our X-ray band, summarizes our searches for FRB counterparts, and describes the calculation of fluence upper limits in the case of non-detections. In Section~\ref{sec:results}, we present the results of these searches followed by comparisons with other studies, exploration of common progenitors with GRBs, and conclude with a discussion of future prospects in the era of forthcoming X-ray missions and radio telescopes.

\section{FRB Data Sample}\label{sec:sample}
For our analysis, we created a catalog by querying the Transient Name Server (TNS), FRBSTATS \citep{frbstats2021}, and FRBCAT \citep{frbcat2016} catalogs for all reported FRBs till September 2024. Table~\ref{tab:frb_sample} provides a summary of our FRB sample. Out of 968 FRBs in our dataset, 723 are non-repeaters, and 245 are repeaters. These detections were made by various observatories, including the Canadian Hydrogen Intensity Mapping Experiment \citep[CHIME;][]{chime2018, chimefirstfrb20219}, the Australian Square Kilometre Array Pathfinder \citep[ASKAP;][]{askap2021}, MeerKAT \citep{meerkat2016}, the Deep Synoptic Array \citep[DSA-110; ][]{dsa10, dsa110}, Parkes \citep{parkes2019}, the upgraded Molonglo Observatory Synthesis Telescope \citep[UTMOST; ][]{utmost2017}, the Green Bank Telescope \citep[GBT; ][]{gbt2009}, and others. This combined catalog is available as \citet{zenodowaratkar2025}.

\begin{table*}[htb]
\tabularfont
\caption{This table contains the break-up of our sample of FRBs divided as non-repeaters and repeaters as detected by various observatories listed, as described in Section~\ref{sec:sample}.}
\label{tab:frb_sample}
\centering
\begin{tabular}{|c|c|c|c|c|c|c|c|}
    \toprule
    Telescope & CHIME & FAST & DSA110 & UTMOST & VLA & Others & Total \\
    \midrule
    Non-Repeaters & 513 & 36 & 56 & 4 & 3 & 111 & 723 \\
    Repeaters & 149 & 65 & 2 & 0 & 1 & 28 & 245 \\
    \bottomrule
    \end{tabular}      
\end{table*}

\subsection{Our FRB sample}\label{subsec:asatsample}
For each of these 968 FRBs, we checked if the source was visible to \asat\ at the moment of the burst, and determined whether the required data were available. When \asat\ passes through the South Atlantic Anomaly (SAA), during which the CZTI's high-voltage supply is switched off for safety reasons, event data are not recorded. Bursts are also missed if the FRB's location is occulted by Earth at the time of arrival (Section~\ref{subsec:arrivaltime}), or if CZTI has data gaps due to slewing or data quality issues. Furthermore, we limited our analysis to FRBs detected after October 6, 2015 (the commencement of CZTI operations), up to August 14, 2023. This selection criteria reduced our sample to 578 FRBs. For these, we searched for burst-like emissions in the CZTI detectors (20--200~keV) as described in the following sections. Details of all 968 FRBs in our sample, including the 578 FRBs with available data, are available at \citet{zenodowaratkar2025}.

\section{Data Analysis}\label{sec:dataanalysis} 
To perform our searches, we first calculate the X-ray arrival times of the FRBs and then run our search pipeline for prompt X-ray counterparts. This involves data reduction, identification of any anomalies, and estimation of flux upper limits in the absence of detections. The steps are outlined in the following subsections.

\subsection{FRB Arrival Times}\label{subsec:arrivaltime} 
As radio waves travel through space, they experience dispersion caused by the intervening plasma between the source and the observer, leading to delays in their arrival compared to shorter wavelength electromagnetic radiation. We calculated the arrival times of all FRBs referenced to an infinite frequency by removing the dispersion delay using following Equation~(\ref{eq:disp_time_delay}).
\begin{equation} 
    \Delta t = 4.15~\mathrm{s} \times \frac{\mathrm{DM}}{\nu^2} \label{eq:disp_time_delay} 
\end{equation}
Here, the dispersion measure (DM) is expressed in $\mathrm{kpc~cm^{-3}}$, and the observation frequency ($\nu$) is given in GHz \citep{lorimerhandbook, petroff2019}. These calculations were performed in the geocentric frame, which was then used for our burst-like searches.

\subsection{CZTI data analyses}\label{subsec:fluxlimit}
We broadly follow the analyses described in \citet{akash2020}, \citet{cift2021}, and \citet{waratkar2024} for reducing the raw \asat\ data to produce cleaned data products and calculate flux upper limits in cases of non-detections of FRBs. The procedure for our analyses is briefly outlined below.

We process the data downlinked from \asat\ using the latest CZTI V3 pipeline\footnote{http://astrosat-ssc.iucaa.in/cztiData}, which employs the \texttt{generalized} event-selection algorithm \citep{ratheesh2021}. This algorithm reduces noise in the CZT detectors, lowering false trigger rates and aiding in the detection of short-duration bursts like FRBs \citep{ratheesh2021}. It includes modules such as \texttt{cztnoisypixclean}, \texttt{cztsuperbunchlean}, \texttt{cztheavybunchclean}, \texttt{cztflickpixclean}, and \texttt{czteventsep}. We generate light curves binned at three timescales — 0.01~s, 0.1~s, and 1~s — within a 20~s search window ($\pm$10~s around the FRB arrival time). We do not search at shorter timescales, as the source would need to be extremely bright to be detectable at such intervals given CZTI's effective area.

As an initial check, we visually inspected these light curves and spectrograms from the CZT detector data (20--200~keV). Following \citet{akash2020}, we estimate the `cut-off count rates', which denote the minimum counts required in a quadrant where the probability of exceeding that count is 10\%. The combined cut-off count rates from all four quadrants give us a false alarm probability (FAP) of $10^{-4}$ for exceeding the threshold. We use detrended data from ten neighboring orbits to produce background histograms, which help estimate the cut-off count rates at the chosen FAP.

We then check for any outliers crossing these thresholds within our search window at the three timescales. In the absence of detections, we estimate the flux and fluence upper limits for X-ray emission in the 20--200~keV range based on the cut-off count rate for each quadrant. Following \citet{waratkar2024}, we exclude any `noisy' quadrant where the cut-off count rate significantly differs from the other three quadrants for the FRBs in question.

\begin{figure*}[!ht]
\centering
\includegraphics[width=\linewidth]{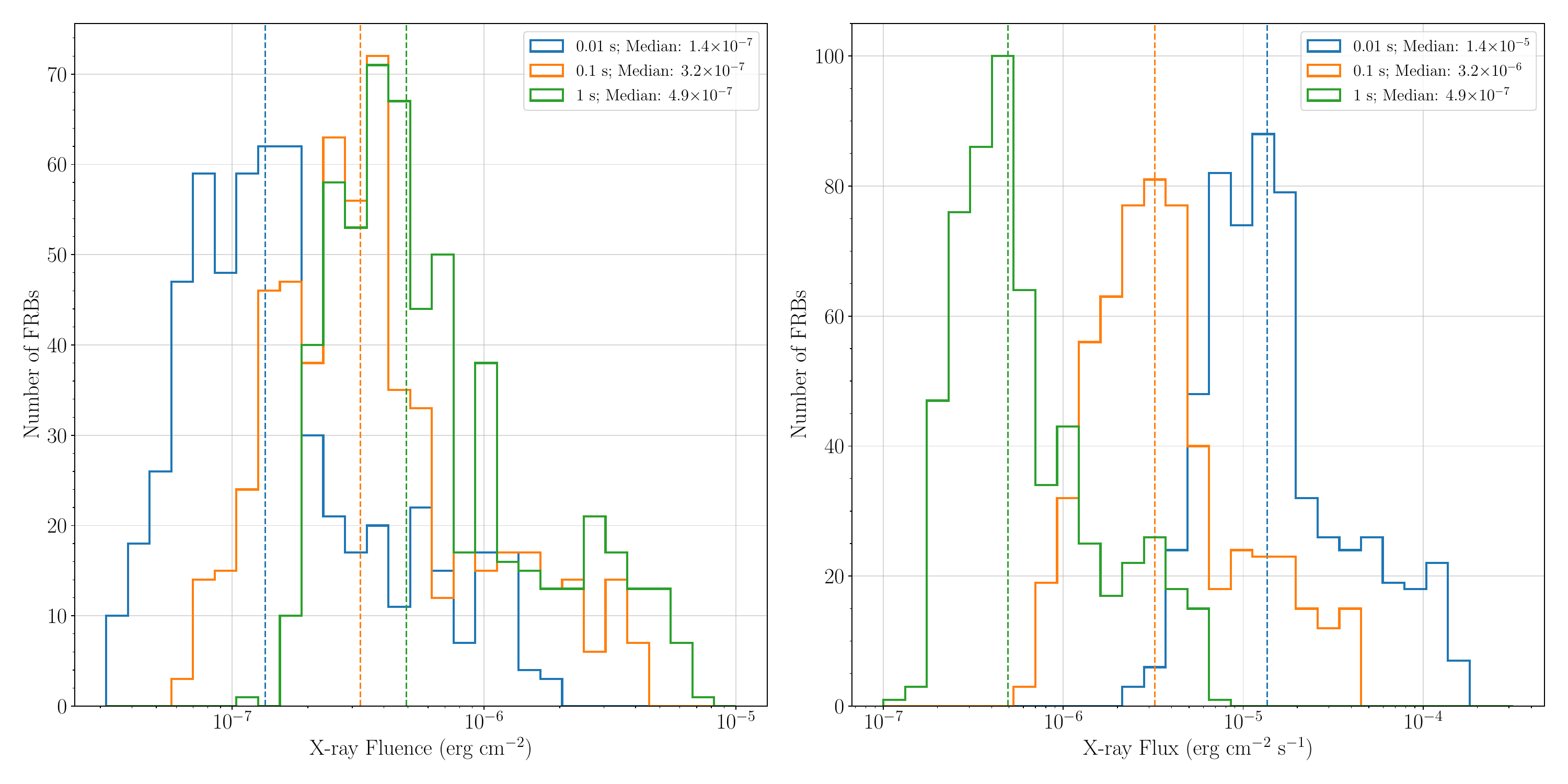}
    \caption{Histogram for the upper limits on X-ray fluences (left) and fluxes (right) for the 20-200~keV band of \czti\ for the FRBs in our sample for all three search timescales --- 0.01~s (blue), 0.1~s (orange), and 1~s (green). We see that the 0.01~s searches give the most constraining fluence limits with a median of $1.4\times10^{-7}$~\eg\, as discussed in Section~\ref{sec:results}. More details about the FRB sample and the limits are available at \citet{zenodowaratkar2025}.}
\label{fig:ul_histograms}
\end{figure*}

We use a pre-calculated satellite response over a 2048--point ``Level~4'' Hierarchical Triangular Mesh (HTM) grid, referred to as the \asat's Mass Model \citep{mate2021}, to calculate CZTI's effective area at the FRB location across different energies. As detailed in \citet{akash2020}, we first assume a power-law spectrum for the FRB (with an index $\mathit{\Gamma} = -1$) and subsequently calculate $\mathit{\Gamma}_\mathrm{max}$, the maximum value of the power-law index possible. We then estimate flux and fluence upper limits for $\mathit{\Gamma}_\mathrm{max}$ at the three search timescales.

\section{Results and inferences from CZTI searches}\label{sec:results}

Out of the 968 FRBs in our sample, we could make inferences for 578 FRBs after excluding those that were occulted by Earth, occurred while \asat\ was in the South Atlantic Anomaly (SAA), or during CZTI data gaps. We found no X-ray counterpart candidates within the 20~s search window for these 578 FRBs. The X-ray flux upper limits for these FRBs in the 20--200keV range were calculated as described in Section~\ref{subsec:fluxlimit}. Figure~\ref{fig:ul_histograms} shows the histogram of the X-ray fluences for all 578 FRBs. The median fluence upper limits are $1.4 \times 10^{-7}$~\eg\ for searches with 0.01~s bins, $3.2\times 10^{-7}$~\eg\ for 0.1~s bins, and $4.9 \times 10^{-7}$~\eg\ for 1~s bins. The 0.01~s searches yield the most stringent limits. We expect the fluence sensitivity to scale with ${t^{1/2}}$, where `$t$' is the ratio of the timescales (10 in this case). However, we observe a slightly different factor of $\sim$2-3 due to non-white noise. The effective area of CZTI varies significantly across the sky, making our flux upper limits dependent on the FRB's relative location to CZTI. All results from these searches are available in \citet{zenodowaratkar2025}, which includes the full list of 578 FRBs in our sample, along with the limits on fluence, flux, ratios, and the computed alphas for all three timescales.

\begin{figure*}[!hp]
\centering
\includegraphics[width=\textwidth]{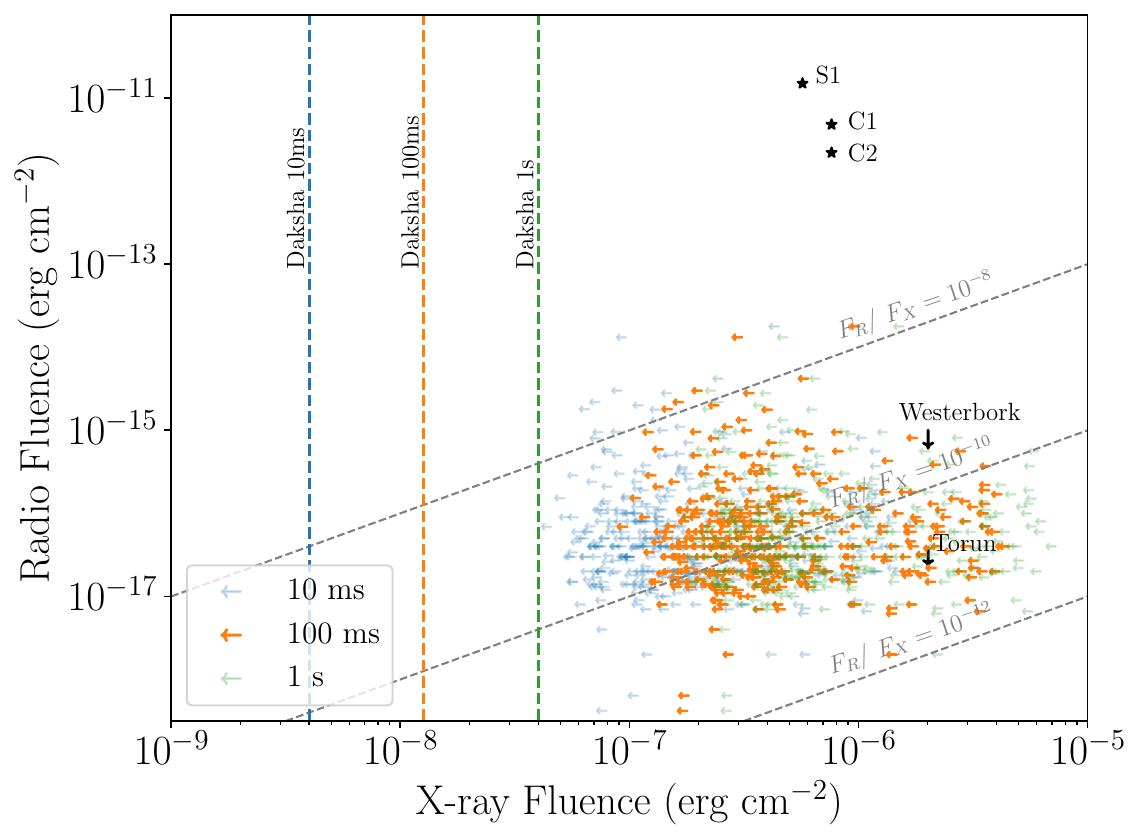}
    \caption{Upper limits (UL) on X-ray fluence for three search timebins --- 0.01~s (blue), 0.1~s (orange), and 1~s (green) --- as a function of radio fluence for FRBs with radio flux measurements. Stars denote the radio and X-ray fluence values for SGR 1935+2154 (S1 -- \citet{sgr12020}; C1 and C2 -- \citet{sgr22020}), while vertical arrows indicate radio upper limits from Westerbork \& Torun \citep{kirstenTorun2021}. The grey dashed lines show the three values of radio to X-ray fluence ratio ($\eta^{-1}$). The dashed vertical lines represents \daksha\ sensitivities for all three time bins \citep{dakshatech2024} which show that \daksha\ will be able to give an order of magnitude deeper limits than \czti\ for the same time binning.}
\label{fig:X-ray_vs_radio_fluence}
\end{figure*}

\subsection{X-ray to Radio fluence ratios}\label{subsec:results_ratios}
Figure~\ref{fig:X-ray_vs_radio_fluence} compares our X-ray upper limits with observed radio fluences for the three different search timescales. The figure presents two cases: one with available radio fluences and another with limits on radio fluences. We also include a comparison with non-detections of fainter radio pulses from SGR~1935+2154 observed by 20--30~m class telescopes at Westerbork and Torun \citep{kirstenTorun2021}, which are consistent with our limits for the extragalactic FRBs in our sample.

\begin{figure*}[ht]
\centering
\includegraphics[width=0.8\linewidth]{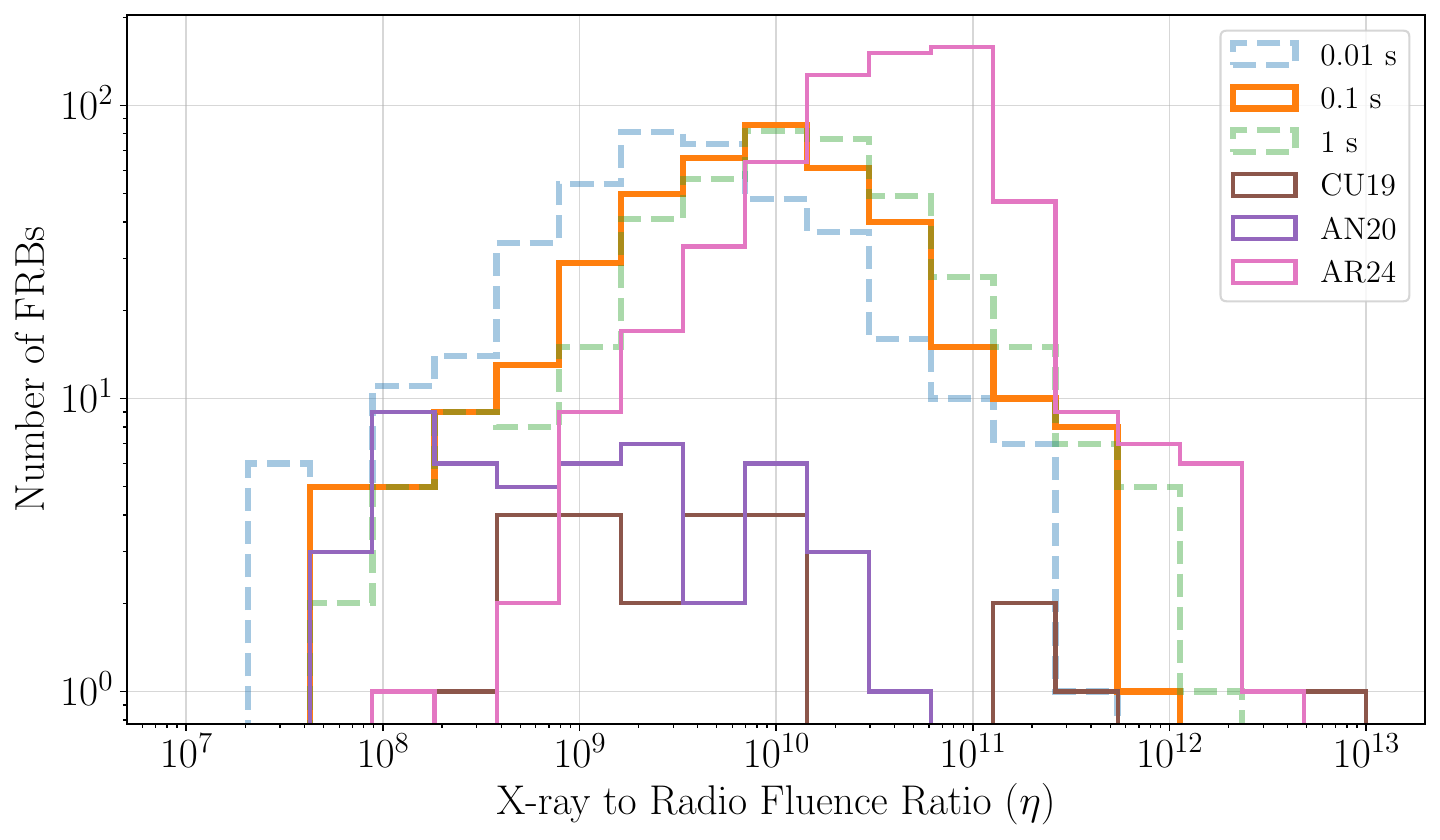}
    \caption{Histogram for the ratios of X-ray to radio fluences for the FRBs in our sample and samples from previous studies \citep{akash2020, cunningham2019,konus2024}. As discussed in Section~\ref{subsec:results_ratios}, we see that the ratios are broadly distributed between $10^{7}$ to $10^{12}$ for all three search timescales --- 0.01~s (blue), 0.1~s (orange), and 1~s (green), which is roughly consistent with the \czti\ limits from the previous searches by \czti\ \citep[violet,][]{akash2020}, \swift-BAT \citep[brown, ][]{cunningham2019}, and KONUS-Wind \citep[pink; ][]{konus2024}.}
\label{fig:ratios_histograms}
\end{figure*}

We further calculate the ratios of our X-ray upper limits to the observed radio fluences. Figure~\ref{fig:ratios_histograms} shows the histogram of these ratios for the 578 FRBs in our sample, with $\eta$ values broadly between $10^{7}$ and $10^{12}$. This is consistent with \citet{akash2020}, but with more samples at higher values of $\eta$: consistent with the fact that surveys are now discovering fainter FRBs, but the X-ray sensitivity has improved by a smaller margin. Our ratios are also consistent with those from \swift-BAT \citep{cunningham2019}, and slightly deeper than more recent KONUS-Wind search \citep[short GRB spectral sample from ][]{konus2024}.

\subsection{Estimating energetics}\label{subsec:results_energetics}

\begin{figure*}
    \centering
    \includegraphics[width=\linewidth]{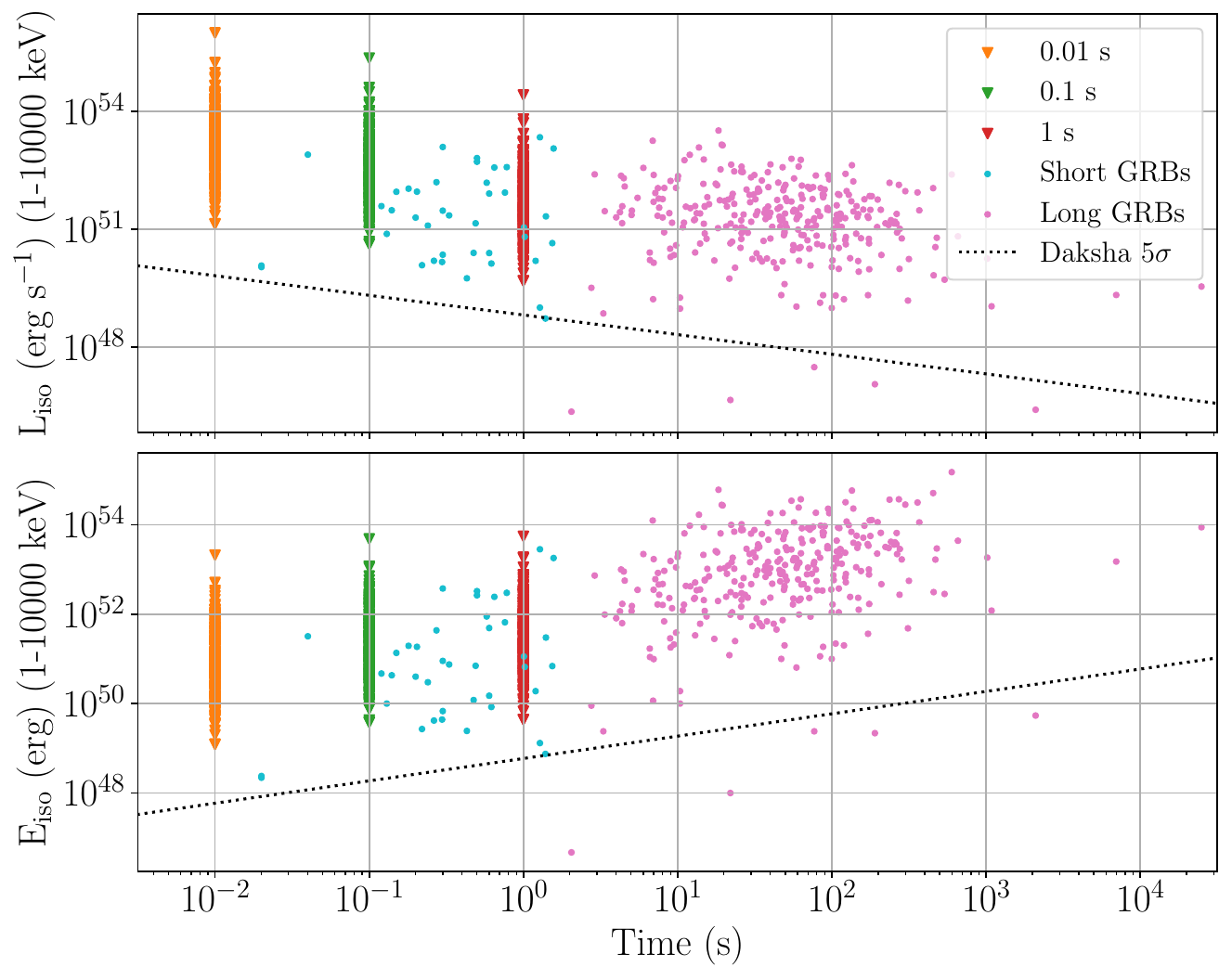}
    \caption{Scatter plot of the limits on isotropic luminosities ($L_\mathrm{iso}$) and isotropic bolometric emission energies ($E_\mathrm{iso}$) for the FRBs in our sample. As discussed in Section~\ref{subsec:results_energetics}, we see that our limits are broadly distributed between $10^{49}$ to $10^{55}$ \es and $10^{49}$ to $10^{53}$ \eg, respectively. We compare these limits with the known luminosities of the gamma-ray bursts (GRBs) in the literature. Our limits imply that the X-ray counterparts of FRBs in our sample are not as energetic as the known GRBs hence further alluding towards different phenomenology for these two populations. It is also possible that the beaming directions for these two emissions could be different. We also show the sensitivity curve of \daksha\ (black dotted) as a function of timescale of the transient in this figure, scaled from redshift of the closest FRB in our sample.}
    \label{fig:energetics_scatter}
\end{figure*}

The DM, a crucial observational property of FRBs, represents the integrated column density of free electrons along the line of sight to the source. It is a by-product of every FRB detection, as the DM value is estimated to correct for the frequency-dependent delays of radio waves. Since the DM contribution from the Milky Way is well understood, it is often used to estimate the distance to the FRB source \citep{ioka2003, inoue2004, zhang2018, pol2019}. For this purpose, we use the \texttt{fruitbat} package \citep{fruitbat2019}, which relies on established DM-redshift relations. Specifically, we employ the Zhang2018 model, known for its reliability across both small and large redshifts \citep{zhangtng2021}, and estimate the luminosity distance from these redshifts using the \texttt{Planck2015} model \citep{planck2015} as implemented in \texttt{astropy.cosmology}. Although about 18 FRBs have known host galaxies and hence accurate known redshifts \citep{tendularhost2017, chatterjeehost2017, vikramhost2019, bhandarihost2022}, we stick to the DM--based distances for all FRBs for uniformity. Switching to host galaxy distance values for these FRBs will not significantly alter our conclusions.

Using these DM-to-distance conversions and our upper limits, we estimate the limits on the energetics of these FRBs, including the high-energy isotropic equivalent luminosities ($L_\mathrm{iso}$) and the isotropic bolometric emission energy ($E_\mathrm{iso}$) in the 1~keV--10~MeV band. These estimates are made in the cosmological rest frame for the standard energy range of 1~keV--10~MeV using the following equations:

\begin{align}
    L_{\mathrm{iso}} & = 4 \pi \, D_{\mathrm{L}}^2 \, F_{\mathrm{UL}} \, k\\
    E_{\mathrm{iso}} & = \frac{4 \pi \, D_{\mathrm{L}}^2 \, k \, F_{\mathrm{UL}}}{1 + z}
\end{align}

where $F_{\mathrm{UL}}$ is the flux upper limit calculated as described above, $D_{\mathrm{L}}$ is the estimated luminosity distance, `$z$' is the estimated redshift of the FRB, and `$k$' is the k-correction factor \citep{bloom2001} to correct for the estimated FRB redshift as well as the energy range of our detectors (20--200~keV), defined as follows:
\begin{equation}
    k =
    \frac{ \int_{1 \, \mathrm{keV} / 1+z}^{10 \, \mathrm{MeV} / 1+z} E \, \frac{dN}{dE}(E) \, dE}{
        \int_{20 \, \mathrm{keV}}^{200 \, \mathrm{keV}} E \, \frac{dN}{dE}(E) \, dE}
\end{equation}

\begin{table*}[!ht]
\tabularfont
    \caption{Key characteristics like the sensitivity, bandwidth, radio field of view (FOV), and expected FRB detection rate for each upcoming radio telescopes, as used for the joint radio and X-ray counterpart rate estimation in the Section~\ref{subsec:results_future}. The FRB detection rate column is presented in units of year$^{-1}$ and represents the radio detected FRB rate.}
\label{tab:radio_details}
\centering    
\begin{tabular}{c|c|c|c|c}
\toprule
Mission & Sensitivity & Bandwidth & Radio FOV & FRB Rate \\
& (Jy ms) & (MHz) & (deg$^2$) & (year$^{-1}$) \\
\midrule
CHIME & 5.0 & 400 & 200 & 530 \\
DSA-2000 & 0.03 & 1300 & 10 & 10000 \\
CHORD & 0.1 & 1200 & 130 & 10000 \\
BURSTT & 5.0 & 400 & 10000 & 100 \\
\bottomrule
\end{tabular}
\end{table*}

\begin{table*}[ht]
\tabularfont
    \caption{Sensitivity and the average sky coverage fraction for the X-ray missions used in the joint radio and X-ray counterpart rate estimation in Section~\ref{subsec:results_future}.}
\label{tab:mission_details}
\centering    
\begin{tabular}{c|c|c}
\toprule
Mission & Sensitivity & Average Sky Coverage \\
& (10$^{-8}$ erg cm$^{-2}$ s$^{-1}$) & \\
\midrule
\czti\ & 52 & 0.7 \\
\fermi-GBM & 20 & 0.7 \\
\swift-BAT & 2 & 0.08 \\
\svom-ECLAIRs & 4 & 0.13 \\
\daksha\ & 4 & 1.0 \\
\bottomrule
\end{tabular}    
\end{table*}

In Figure~\ref{fig:energetics_scatter}, we show the $L_\mathrm{iso}$ and $E_\mathrm{iso}$ limits estimated from our \czti\ upper limits using the methodology described above. From these plots, it is evident that our $L_\mathrm{iso}$ limits are broadly distributed between $10^{49}$ and $10^{55}$ \es, while the $E_\mathrm{iso}$ limits span $10^{49}$ to $10^{53}$ \eg. We can compare these limits with the known luminosities of GRBs \citep{amati2006, berger2014, grbcat2023}, which have $L_\mathrm{iso}$ and $E_\mathrm{iso}$ values ranging from $10^{49}$ to $10^{54}$ \es and $10^{50}$ to $10^{55}$ \eg, respectively, higher than our limits for many events. We also compare the timescales of these two populations of transients, further dividing the GRB sample into short-duration and long-duration GRBs \citep{grbclass1993}. Our limits suggest that the progenitors are unlikely to be either short GRBs (with durations less than 2~s) or long GRBs (with durations greater than 2~s) --- with the exception of low-luminosity GRBs. If lower luminosity GRBs such as the short-GRB counterpart to GW170817 \citep{bns12017}, with an isotropic luminosity ($L_\mathrm{iso} = 1.6 \times 10^{46}$ \es\ and $E_\mathrm{iso} = 5.3 \times 10^{46}$ \eg) \citep{bns12017, bns22017} were indeed associated with FRBs, our searches would not have detected them. A similar conclusion was also reached for X-ray counterpart searches in INSIGHT-HXMT data \citep{hxmt2020}. 

Future missions like \daksha\ \citep{dakshatech2024}, with its all-sky coverage, wide energy band, and an order of magnitude higher sensitivity compared to CZTI, will be crucial for detecting more low-luminosity GRBs \citep{dakshascience2024}. We use the distance to the closest FRB in our sample (FRB~20181030B), and show the sensitivity of \daksha\ at different search timescales in Figure~\ref{fig:energetics_scatter}. We see that such a high--sensitivity mission can search for significantly fainter hard X-ray counterparts to FRBs.

Lastly, we note that $E_\mathrm{iso}$ of the SGR~1935+2145 galactic magnetar is $\sim 10^{39}$~\eg\ \citep{integralsgr2020}, several orders of magnitude lower than the luminosity constraints from any of these works. The X-ray to radio fluence ratio for this source is $\sim 10^{5}$ \citep{integralsgr2020}, which is at least three orders of magnitude lower than the ratios for any of the extragalactic FRBs in our sample.

\subsection{Future Prospects for Joint Radio and X-ray Detections}\label{subsec:results_future} 
Given the lack of X-ray detections from extragalactic FRBs, the most promising approach to detect or place meaningful constraints on coincident X-ray signatures is to develop more sensitive X-ray monitors or focus on detecting brighter, nearby FRBs. To explore the future of FRB detections, we turn to upcoming observatories like BURSTT \citep{burstt2022}, CHIME \citep{chime2018, chime2022}, CHORD \citep{chord2020}, and DSA-2000 \citep{dsa2000}\footnote{\url{https://www.deepsynoptic.org/overview}}. Each of these observatories has unique advantages and is expected to enable numerous bright FRB detections, as shown in Table~\ref{tab:radio_details}. We also estimate the performance of existing and future X-ray monitors, including \fermi-GBM \citep{fermi2009}, \swift-BAT \citep{swift2004,bat2005}, \svom-ECLAIRs \citep{svom2015,eclairs2014}, and \daksha\ \citep{dakshatech2024}, as shown in Table~\ref{tab:mission_details}.

\begin{figure*}[!ht]
\centering
\includegraphics[width=\linewidth]{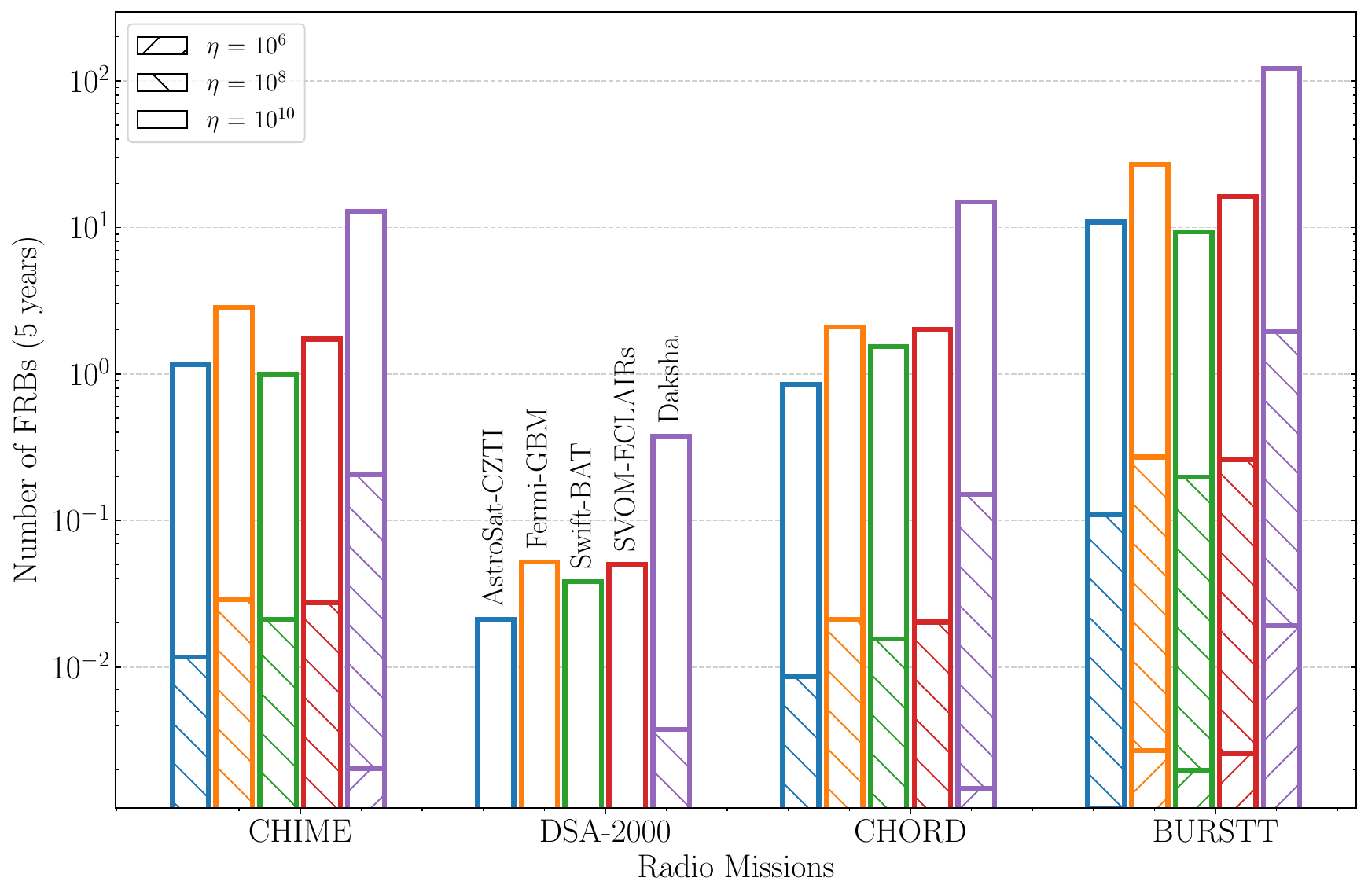}
    \caption{Number of joint X-ray and radio detections of FRBs by various instrument pairs, for different assumed values of the X-ray to Radio fluence ratios --- \czti, \fermi-GBM, \swift-BAT, \svom-ECLAIRs, and \daksha\ --- for a 5-year operation period. We assume three different X-ray to radio fluence ratios ($\eta$) of $10^6$, $10^{8}$, and $10^{10}$ as discussed in Section~\ref{subsec:results_future}. Large field-of-view missions like \fermi-GBM, \czti, and \daksha\ have the highest potential for joint X-ray and radio detections of FRBs when paired with wide-field radio telescopes such as BURSTT. For highly sensitive radio instruments like DSA-2000, CHIME, and CHORD, the majority of detected FRBs are likely too faint for concurrent X-ray detection, unless the X-ray to radio emission ratio ($\eta$) is high. The combination of \daksha\ and BURSTT offers the most promising opportunity for detecting joint X-ray and radio counterparts, due to \daksha's high sensitivity in X-rays and BURSTT's large field of view in radio.}
\label{fig:future_frbs}
\end{figure*}

We consider the FRB population detected by CHIME \citep{chimecat12021}, where the rates of FRB detection as a function of fluence follow a power-law distribution with $\Gamma = -2$ \citep{mcgregor2024}. We assume that this power-law distribution extends beyond the sensitivity limit of CHIME and holds true for other telescopes, such as CHORD and DSA-2000, which have better sensitivity. We normalize the population distribution to match the predicted FRB population from these telescopes \citep{mcgregor2024, burstt2022}. 

For a burst to be detected by a given X-ray mission (\daksha, for instance), its X-ray fluence must be higher than the sensitivity limit of that mission. We use three nominal values of $\eta$ ($10^6$, $10^8$, and $10^{10}$) to calculate the corresponding radio fluence limit ($F_\mathrm{radio,min}$) for such bursts. Next, we use the sensitivity of various radio surveys, and calculate the number ($N_\mathrm{joint,max}$) of FRBs detectable by them that are brighter than $F_\mathrm{radio,min}$. 

$N_\mathrm{joint,max}$ is the expected number of joint detections if both the X-ray and the radio telescopes continuously observed the entire sky. The ability of an X-ray satellite to place constraints on an FRB detected by a radio telescope depends on the overlap of their instantaneous fields of view, affecting the rate of joint detections over time. We account for this effect by multiplying sky fractions coverage of both the radio telescope and the X-ray mission with the above-estimated rates of FRBs whose counterparts would be detected by the X-ray missions\footnote{Note that this assumes that the X-ray telescope and radio telescope pointing is completely independent of each other.}. For \swift-BAT and \svom-ECLAIRs, while their FoV is 0.11 and 0.16 respectively, we also account for the sky fraction (30\%) occulted by Earth.

Figure~\ref{fig:future_frbs} shows these joint detection rates for a 5-year operational period. We observe that large field-of-view missions like \fermi-GBM, \czti, and \daksha complement large field-of-view radio telescopes like BURSTT, leading to the highest number of joint X-ray and radio detections of FRBs. While DSA-2000 detects a large number of FRBs with its high sensitivity, most are faint and do not have detectable X-ray counterparts. However, if $\eta$ is high, the putative X-ray counterparts will be bright enough to be detected by X-ray missions, increasing the joint rates for DSA-2000. A similar effect is seen for CHIME and CHORD, which have similar fields of view: CHORD will detect more FRBs, but most of them will be too faint to be seen by X-ray missions, unless $\eta$ is high. Note that we do not account for the energy band of the X-ray missions in this analysis, which could affect the number of joint detections.

\section{Conclusion}\label{sec:conclusion}
We undertook a systematic search for prompt X-ray counterparts of FRBs in \czti\ data. We selected a sample of 978 FRBs detected from the launch of \asat\ in October 2015 to August 2023. A total of 400 FRBs were excluded from our sample due to unavailable data, which was either due to the satellite passing over the South Atlantic Anomaly (SAA), CZTI data gaps, or the FRB being occulted by Earth, resulting in a sample of 578 FRBs for which our searches were possible. While our searches yielded no credible X-ray counterparts, we were able to place meaningful upper limits on the X-ray flux and fluence in the 20--200~keV range, and further estimate the X-ray to radio fluence ratios for these FRBs. These ratios are in agreement with those from previous searches.

Using DM-to-redshift conversions to estimate the luminosity distances to the FRBs, we constrained the source isotropic equivalent luminosities to be $ <10^{49}$ to $10^{55}$~\es, with corresponding $E_\mathrm{iso} < 10^{49}$ to $10^{53}$~erg. The lower ends of these values are less than luminosities of typical short and long GRBs, disfavouring the associations of FRBs with these sources. However, we cannot rule out the possibility of an association with low-luminosity GRBs.

We explored the future prospects for joint radio and X-ray detections of FRBs using upcoming radio and X-ray missions, considering three scenarios with varying X-ray to radio fluence ratios. We found that large field-of-view missions like \fermi-GBM, \czti, and \daksha\ complement large field-of-view radio telescopes like BURSTT, resulting in the highest number of joint X-ray and radio detections of FRBs. Among the telescopes considered in this study, the largest number of joint detections are expected for \daksha\ with its  high sensitivity and all-sky coverage, combined with the large field of view of BURSTT. Other telescopes like CHORD, CHIME, and DSA-2000 are likely to result in joint detections mainly for larger values of $\eta$.

A deeper understanding of these enigmatic sources, their emission mechanisms, and their progenitors hinges on detecting more FRBs and their counterparts across different wavelengths. This requires a joint effort from both the radio and X-ray communities, with a focus on identifying bright, well-localized FRBs that potentially offer the best opportunity to detect their counterparts at other wavelengths. Rapid public dissemination of results is crucial for facilitating timely observations (e.g., \citet{frb-voe2025}). Public alerts allow the community to check for temporal and spatial coincidences between FRBs and other transients, helping prioritize the most promising candidates for follow-up. For instance, the Rapid, On-Source VOEvent Coincidence Monitor \citep[RAVEN; ][]{ravenurban, ravencho, ravenbrandon} identifies GRBs coincident with gravitational wave alerts for immediate follow-up—a similar system for FRBs would be highly beneficial. Additionally, coordinated multi-wavelength campaigns targeting repeating FRBs could probe fainter bursts and their counterparts, offering deeper insights into their origins.

\section*{Acknowledgements}
We thank Kritti Sharma, Vikram Rentala, A. R. Rao, and Gulab Dewangan for fruitful discussions. We also thank Akash Anumarlapudi and Vedant Shenoy for their contributions to the initial setting up of the codes that were then further developed by the current team. Lastly, we thank the CZTI Interface for Fast Transients (CIFT) Team at IIT Bombay and the Payloads Operations Center (POC) team at IUCAA, Pune, for their valuable contributions to the regular transient searches in the \czti\ data.

CZTI is built by a consortium of Institutes across India. The Tata Institute of Fundamental Research, Mumbai, led the effort with instrument design and development. Vikram Sarabhai Space Centre (VSSC) at Thiruvananthapuram provided the electronic design, assembly, and testing. ISRO Satellite Centre (ISAC) at Bengaluru provided the mechanical design, quality consultation, and project management. The Inter-University Center for Astronomy and Astrophysics (IUCAA) at Pune did the Coded Mask design, instrument calibration, and the POC. The Space Application Centre (SAC) at Ahmedabad provided the analysis software. The Physical Research Laboratory (PRL) Ahmedabad provided the polarization detection algorithm and ground calibration. A vast number of industries participated in the fabrication, and the university sector pitched in by participating in the testing and evaluation of the payload. The Indian Space Research Organisation (ISRO) funded, managed, and facilitated the project.

Software: This work utilized various software, including \texttt{Python} \citep{python3}, \texttt{Astropy} \citep{astropy}, \texttt{NumPy} \citep{numpy}, and \texttt{Matplotlib} \citep{matplotlib}. Some of the results of this paper have been derived using the \texttt{FRUITBAT} package. This research has made use of NASA's Astrophysics    Data System. We also acknowledge the use of the Zenodo repository for sharing our data products \citep{zenodo}.

\bibliographystyle{apj}
\bibliography{bibiliography}

\end{document}